\documentclass[aps,prl,twocolumn,superscriptaddress]{revtex4-1}
\usepackage{graphicx}
\begin{document}

\title{Coexistence of Ferroelectric-like Polarization and Dirac-like Surface State in TaNiTe$_5$}
\author{Yunlong Li}
\email{equally contributed}
\author{Zhao Ran}
\email{equally contributed}
\author{Chaozhi Huang}
\affiliation{Key Laboratory of Artificial Structures and Quantum Control (Ministry of Education), Shenyang National Laboratory for Materials Science, School of Physics and Astronomy, Shanghai Jiao Tong University, Shanghai 200240, China}
\author{Guanyong Wang}
\affiliation{Shenzhen Institute for Quantum Science and Engineering, Southern University of Science and Technology, Shenzhen 518055, China}
\author{Peiyue Shen}
\author{Haili Huang}
\affiliation{Key Laboratory of Artificial Structures and Quantum Control (Ministry of Education), Shenyang National Laboratory for Materials Science, School of Physics and Astronomy, Shanghai Jiao Tong University, Shanghai 200240, China}
\author{Chunqiang Xu}
\affiliation{School of Physics and Key Laboratory of MEMS of the Ministry of Education, Southeast University, Nanjing 211189, China}
\author{Yi Liu}
\affiliation{Department of Applied Physics, Zhejiang University of Technology, Hangzhou 310023, China}
\author{Wenhe Jiao}
\affiliation{Ningbo Institute of Materials Technology and Engineering, Chinese Academy of Sciences, Ningbo 315201, China}
\author{Wenxiang Jiang}
\author{Jiayuan Hu}
\author{Gucheng Zhu}
\author{Chenhang Xu}
\author{Qi Lu}
\author{Guohua Wang}
\author{Qiang Jing}
\author{Shiyong Wang}
\author{Zhiwen Shi}
\affiliation{Key Laboratory of Artificial Structures and Quantum Control (Ministry of Education), Shenyang National Laboratory for Materials Science, School of Physics and Astronomy, Shanghai Jiao Tong University, Shanghai 200240, China}
\author{Jinfeng Jia}
\affiliation{Key Laboratory of Artificial Structures and Quantum Control (Ministry of Education), Shenyang National Laboratory for Materials Science, School of Physics and Astronomy, Shanghai Jiao Tong University, Shanghai 200240, China}
\affiliation{Tsung-Dao Lee Institute, Shanghai Jiao Tong University, Shanghai 200240, China}
\author{Xiaofeng Xu}
\affiliation{Department of Applied Physics, Zhejiang University of Technology, Hangzhou 310023, China}
\author{Wentao Zhang}
\email{wentao@sjtu.edu.cn}
\affiliation{Key Laboratory of Artificial Structures and Quantum Control (Ministry of Education), Shenyang National Laboratory for Materials Science, School of Physics and Astronomy, Shanghai Jiao Tong University, Shanghai 200240, China}
\author{Weidong Luo}
\email{wdluo@sjtu.edu.cn}
\affiliation{Key Laboratory of Artificial Structures and Quantum Control (Ministry of Education), Shenyang National Laboratory for Materials Science, School of Physics and Astronomy, Shanghai Jiao Tong University, Shanghai 200240, China}
\author{Dong Qian}
\email{dqian@sjtu.edu.cn}
\affiliation{Key Laboratory of Artificial Structures and Quantum Control (Ministry of Education), Shenyang National Laboratory for Materials Science, School of Physics and Astronomy, Shanghai Jiao Tong University, Shanghai 200240, China}
\affiliation{Tsung-Dao Lee Institute, Shanghai Jiao Tong University, Shanghai 200240, China}

\date{\today}

\begin{abstract}

By combining angle-resolved photoemission spectroscopy (ARPES), scanning tunnelling microscopy (STM), piezoresponse force microscopy (PFM) and first-principles calculations, we have studied the low-energy band structure, atomic structure and charge polarization on the surface of a topological semimetal candidate TaNiTe$_5$. Dirac-like surface states were observed on the (010) surface by ARPES, consistent with the first-principles calculations. On the other hand, PFM reveals a switchable ferroelectric-like polarization on the same surface. We propose that the noncentrosymmetric surface reconstruction observed by STM could be the origin of the observed ferroelectric-like state in this novel material. Our findings provide a new platform with the coexistence of ferroelectric-like surface charge distribution and novel surface states.

\end{abstract}


\maketitle

Ferroelectric materials have a spontaneous polarization with the noncentrosymmetric crystal structure\cite{Ferro_Book}. Conventionally, ferroelectricity exists in insulators or semiconductors rather than in metals due to the screening effect of the conduction electrons. In 1960s, Anderson and Blount proposed the possibility of a ferroelectric metal with inversion asymmetric crystal structure in the metallic state\cite{Anderson}. Since then, many efforts have been made to explore the ferroelectricity in metallic systems. Recently, the centrosymmetric to noncentrosymmetric structural transition has been observed in Cd$_2$Re$_2$O$_7$ at 200 K\cite{Cd2Re2O7} and LiOsO$_3$ at 140 K\cite{LiOsO3} with metallic transport behaviour, which possibly sustain ferroelectricity in metals. Polar domains and strain induced ferroelastic switching were observed in bulk polar metal Ca$_3$Ru$_2$O$_7$ lately\cite{Ca3Ru2O7}. Other strategies have also been used to obtain the ferroelectric metals, such as doping the ferroelectric insulators\cite{dopeBTO1,dopeBTO2} and engineering interface based polar metals in oxide heterostructures\cite{dopePTNO}. In very thin polar metal, like bilayer and trilayer WTe$_2$, where an electric field can penetrate, ferroelectric switching has been observed\cite{WTe2transport}. Interestingly, WTe$_2$ has also been demonstrated to be a topological semimetal\cite{WS_WTe2_1,WS_WTe2_2,WS_WTe2_3}. Therefore, WTe$_2$ becomes a unique system hosting both ferroelectric and topological phases. The combination of these two quantum states of matter has intriguing potential for future applications.

In principle, a possible land where to seek the coexistence of ferroelectric and topological state is the surface of topological materials. Topological materials possess novel surface states, and possible reconstruction of the surface atoms might cause noncentrosymmetric surface layers supporting ferroelectric-like polarization.

In this work, by combining ARPES, STM and PFM measurements, we observed the ferroelectric-like polarization together with novel surface states on the (010) surface of a topological semimetal TaNiTe$_5$. Although the bulk TaNiTe$_5$ is centrosymmetric and thus non-polar, we found a noncentrosymmetric reconstruction of the surface atoms, which could play an essential role in forming the ferroelectric-like polarization on the surface. Experimental detected Dirac-like surface state matches the calculations based on the bulk crystal structure very well. Our findings provide a new interesting platform with the coexistence of ferroelectric-like state and topological state.

ARPES measurements were performed at ARPES beamline in National Synchrotron Radiation Laboratory (Hefei) and at BL-03U/09U in Shanghai Synchrotron Radiation Facility. The time-resolved ARPES (tr-ARPES) measurements were carried out to detect unoccupied state. STM measurements were carried out in a low-temperature STM system at 77 K. The PFM measurements were performed on a Cypher S atomic force microscope (Oxford Instruments) in air at room temperature. Tips used in PFM measurements are commercial SCM-PIT-75 (NanoWorld). The PFM images were acquired at an AC imaging bias of 1 V and a frequency of approximately 300 kHz at the resonant peak. Density functional theory (DFT) calculations were performed using VASP\cite{VASP1,VASP2}. The plane-waves basis with an energy cutoff of 500 eV was used and the exchange-correlation functional was given by GGA-PBE\cite{GGA-PBE}. The bulk crystal structure was relaxed until the residual force on each atom is smaller than 0.001 eV/\AA. A 9 $\times$ 3 $\times$ 3 k-grid was used\cite{Monkhorst-Pack}. We employed the Wannier90 code\cite{Wannier90} to construct the tight-binding Hamiltonian from the VASP self-consistent calculations and then obtain the surface states based on a Green function scheme\cite{Green_function,WannierTools}.

TaNiTe$_5$ has a layered orthorhombic structure (space group $Cmcm$) with lattice parameters $a$ = 3.667 \AA, $b$ = 13.172 \AA, $c$ = 15.142 \AA\cite{Structure1}. Shown in Fig. 1(a), the crystal structure consists of bicapped trigonal
prismatic Ta chains and octahedral Ni chains along the $a$-axis. The two chains face-share with each other along the $c$-axis forming layers stacking along the $b$-axis. The bulk Brillouin zone (BZ) and its surface projection to the (010) surface is shown in Fig. 1(b). The $x/y/z$ direction in Fig. 1(b) corresponds to the $a/b/c$-axis in Fig. 1(a). Figure 1(c) presents the calculated bulk bands near the Fermi energy (E$_F$). Both hole-like and electron-like bands cross the E$_F$, so TaNiTe$_5$ is metallic. Band dispersion along the $b$ and $c$-axis is much weaker than that along the $a$-axis, indicating TaNiTe$_5$ has quasi-one-dimensional property to some extent. Previous calculations and transport measurements suggested TaNiTe$_5$ is a topological semimetal\cite{TNT1}.

\begin{figure}[t]
\centering
\includegraphics[width=8cm]{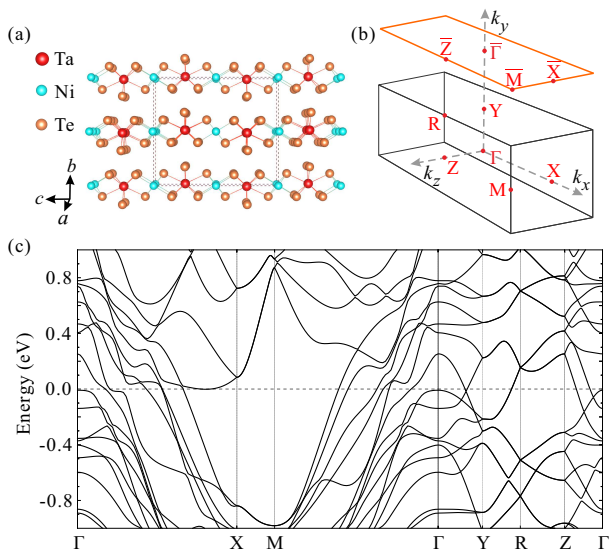}
\caption{(a) Crystal structure of TaNiTe$_5$. Dashed lines indicate the unit cell. (b) Bulk Brillouin zone and its surface projection to the (010) surface. (c) Calculated bulk band structure of TaNiTe$_5$.}
\end{figure}

\begin{figure*}[]
\centering
\includegraphics[width=16cm]{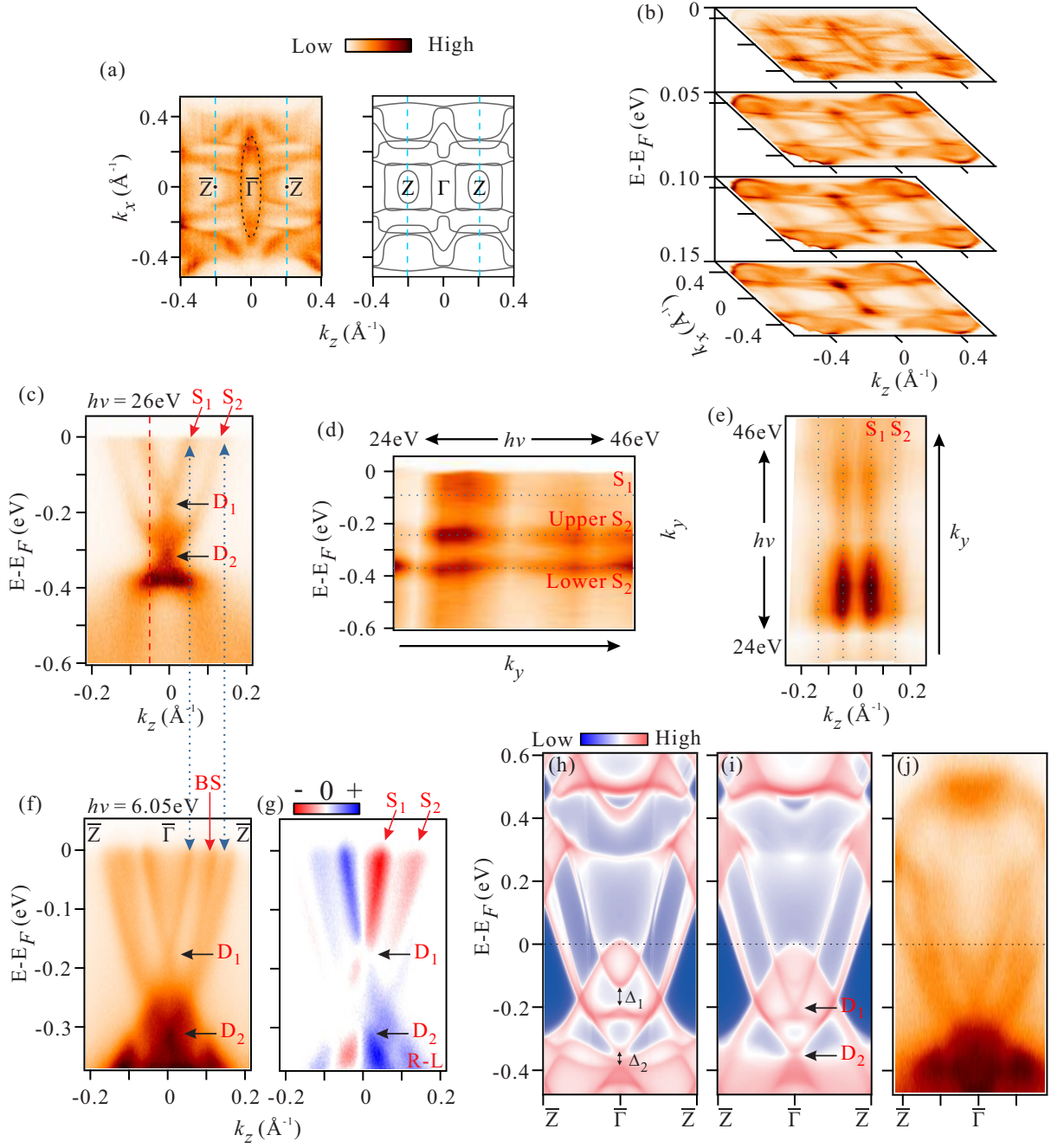}
\caption{(a) FS mapping of TaNiTe$_5$ using h$\nu$ = 26 eV photons and the calculated bulk FS in the $\Gamma ZMX$ plane. Blue dashed lines mark the surface BZ boundary. Black dashed line markes the elliptical FS that does not exist in the calculated bulk FS. (b) Constant energy contours at different binding energies. The elliptical FS becomes smaller at higher binding energy. (c) ARPES band dispersion along $\bar{Z}-\bar{\Gamma}-\bar{Z}$ direction crossing the $\bar{\Gamma}$ point. (d) Band dispersion along $k_y$ direction. (e) FS mapping in ${\Gamma}YRZ$ plane. There is no obvious $k_y$ dispersion. (f) ARPES spectra covering the same momentum position as in (c) using h$\nu$ = 6.05 eV photons. (g) Intensity difference between ARPES spectra using right and left circular polarization incident lights. Clear circular dichroism effect is observed. (h) Calculated spectral function of bulk bands projection to the (010) surface. (i) Calculated spectra function of the surface states and the bulk states projection on the (010) surface. Two Dirac-like surface states appear in the bulk band gap. (j) Tr-ARPES spectra at delay time of about 0.1 ps. Unoccupied states were observed.}
\end{figure*}

We carried out ARPES experiments on the cleaved (010) surface (\textit{ac}-plane). Figure 2(a) shows the Fermi surface (FS) mapping (left) using incident photon of 26 eV and the calculated FS (right) in the $\Gamma ZMX$ plane. Consistent with the calculations, several open FS sheets along the k$_z$ direction and a large FS pocket surrounding $\bar{Z}$ point were observed. According to the crystal structure in Fig. 1(a), k$_z$ ($c$-axis) is the in-plane direction perpendicular to the $a$-axis with chain-like structure, therefore it presents quasi-one-dimensional electronic property to a certain extent. The most obvious deviation between the experiments and the calculations is the existence of an elliptical Fermi pocket (marked by black dashed lines in Fig. 2(a)) surrounding $\bar{\Gamma}$ point. Constant energy contours at different binding energies are shown in Fig. 2(b). The elliptical pocket becomes smaller at higher binding energy, indicating it is an electron-like pocket.

ARPES spectra along the $k_z$ direction crossing $\bar{\Gamma}$ point is presented in Fig. 2(c). Two bands disperse from the zone center towards the zone boundary and cross the E$_F$ at $k_z\sim$ 0.05 \AA$^{-1}$ and 0.13 \AA$^{-1}$. We label these two bands as band ``S$_1$'' and band ``S$_2$''. Pointed by black arrows, two band crossing points (``D$_1$'' and ``D$_2$'') are observed. Photon energy dependent measurements were performed to check the surface/bulk origin of these two bands. Figure 2(d) shows the band dispersion along the $k_y$ (normal direction to the cleaving surface) with fixed in-plane momentum k$_z$ (indicated by red dashed line in Fig. 2(c)). No obvious $k_y$ dispersion is observed for both bands. Furthermore, Fig. 2(e) shows FS mapping in ${\Gamma}YRZ$ plane ($k_{yz}$ plane). FSs of S$_1$ and S$_2$ bands are just vertical lines along the k$_y$ direction, which also indicates no band dispersion along k$_y$. Therefore, we can conclude that bands S$_1$ and S$_2$ are two-dimensional surface states.

High-resolution band dispersions along $\bar{Z}-\bar{\Gamma}-\bar{Z}$ direction was further studied by ARPES using 6.05 eV laser in Fig. 2(f). Besides the S$_1$ and S$_2$ bands, we detected another band (labelled as band ``BS''). BS band is very likely a bulk band due to the higher bulk sensitivity for 6.05 eV photons comparing with 26 eV photons\cite{probe_depth}. It is more clear that band S$_1$ has a linear-like band dispersion and crosses at the D$_1$ point, therefore D$_1$ is a Dirac point. In contrast to Fig. 2(c), D$_2$ point is buried in some fuzzy features although it is still resolvable. Those fuzzy features should originate from bulk band contributions. The possible spin information of the observed bands are checked by circular dichroism (CD) effect of ARPES\cite{CD}. Intensity difference between ARPES spectra using right and left circularly polarized incident lights is plotted in Fig. 2(g) and two bands are resolved. According to the momentum position, they are bands S$_1$ and S$_2$. Those two bands show clear CD effect that reverses above and below the crossing points of D$_1$ and D$_2$, implying the spin-momentum locking property of the surface state. Band BS is not resolved in Fig. 2(g), consistent with its possible bulk origin, since bulk bands in TaNiTe$_5$ is spin degenerate. It is worth noting that S$_1$ and S$_2$ show the similar CD effect, which excludes that they are a pair of Rashba splitting bands. S$_1$ and S$_2$ are indeed two different bands.

Surface bands S$_1$ and S$_2$ are nicely reproduced by our DFT calculations. Figure 2(h) shows the calculated spectral function of bulk bands projection along $\bar{Z}-\bar{\Gamma}-\bar{Z}$ direction. Obviously, no S$_1$ band and Dirac point D$_1$ exists in bulk states. Instead of D$_1$ point, there is a bulk gap $\Delta_1$ of $\sim$ 0.1 eV at $\bar\Gamma$ point. There is also no D$_2$ point at $\sim$ 0.3 eV below E$_F$, but a small energy gap $\Delta_2$ of $\sim$ 0.05 eV at $\bar\Gamma$ point. Figure 2(i) presents the calculated spectral function of the surface states and the bulk states projection on the (010) surface. Pointed by black arrows, two surface bands with crossing points appear in the bulk band gap. Compared with Fig. 2(j), the experimental band dispersion matches the calculated surface bands very well. From Fig. 2(i), we know that most of the surface band S$_2$ nearly overlaps with the bulk states away from the Dirac point D$_2$. Unoccupied states above E$_F$ were also explored by tr-ARPES measurements, presented in Fig. 2(j). Again, the nice agreement between the experiments and the calculations in Fig. 2(h) above the E$_F$ is observed.

\begin{figure*}[t]
\centering
\includegraphics[width=16cm]{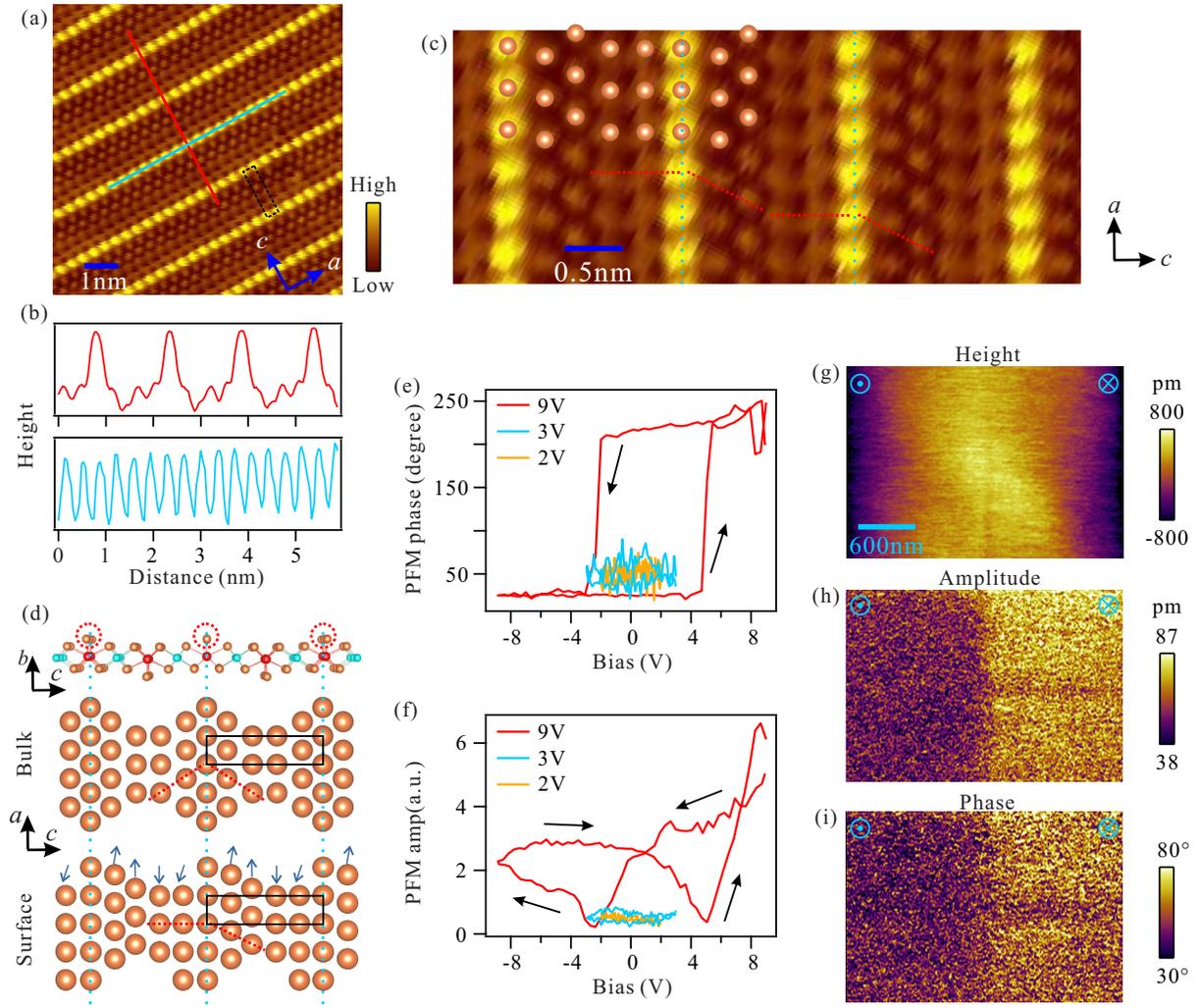}
\caption{(a) Atomically resolved STM image (Bias=3.55 mV, current=100 pA) on the (010) surface of TaNiTe$_5$. Black rectangle marks the surface unit cell. (b) Line profiles based on the red and blue lines in (a). (c) Zoomed-in STM image of (a) to reveal the position of Te atoms. Orange circles mark the Te atoms. Blue dotted lines indicate the highest Te atoms. Red dotted lines are guide to the eye to show the loss of inversion symmetry. (d) Top, side view of a Te-Ta-Ni-Te layer. Middle, top view of Te atoms on the (010) surface based on the bulk lattice structure. The bulk structure possesses inversion symmetry in the plane. Bottom, top view of Te atoms on the (010) surface based on the STM measurements. No inversion symmetry in the plane. Black rectangles mark the in-plane unit cell. Blue arrows indicate the direction of the movement of atoms comparing to the bulk structure. PMF phase (e) and amplitude (f) as a function of bias voltage under the different maximum bias voltages. Hysteresis loop and butterfly-shape loop were observed above the threshold voltage. (g) Surface topograph (height), (h) PFM amplitude and (i) PFM phase images from a poled TaNiTe$_5$ area. The area was poled by applying DC bias to the tip (left : -8 V, right : +8 V).}
\end{figure*}

Atomically resolved STM image on the (010) surface is shown in Fig. 3(a). According to the crystal structure, the cleaving surface should be Te atoms terminated due to the weak inter-layer interaction between Te atoms. Chain-like structure along the $a$-axis is visualized, consistent with the crystal structure in Fig. 1(a). The brightest features are the topmost Te atoms on the Ta chains. The surface unit cell indicated by the black rectangle in Fig. 3(a) is the same as the bulk. Extracted from the line profiles in Fig. 3(b), lattice constants along the $a$ and $c$-axis are $a_{sur}$= 3.6$\pm$0.1 \AA \ and $c_{sur}$= 14.9$\pm$0.1 \AA \, that are slightly smaller than the bulk value ($a$=3.667 \AA \ and $c$=15.142 \AA) within the experimental uncertainty. The positions of surface Te atoms are presented in Fig. 3(c). Orange circles mark the Te atoms. Interestingly, although the surface lattice constants and surface unit cell are very similar to the bulk, the distribution of surface Te atoms is quite different. According to the bulk crystal structure (Fig. 3(d)), Te atoms in the $ac$-plane possess inversion symmetry. However, our experiments indicate surface Te atoms do not have inversion symmetry. In Fig. 3(c), it is clear that the atoms on the left and right side of the topmost Te chain (blue dotted line marks the Te chain) lose mirror symmetry. We summarize the top view of two kinds of configurations of Te atoms on the (010) surface in Fig. 3(d). They have identical rectangular lattice, but very different symmetry.

In a metallic system, inversion asymmetry is necessary for the ferroelectricity. Although we only obtained the accurate information about the in-plane position of the surface Te atoms from STM measurements, all atoms in the top Te-Ta-Ni-Te layer should rearrange their position both in-plane and out of plane in principle. Therefore, if there is charge polarization on the surface, the polarization direction could have out-of-plane component. To probe the possible ferroelectricity in TaNiTe$_5$, we carried out PFM experiments at room temperature. PFM is sensitive to the out-of-plane charge polarization. The PFM phase and amplitude as a function of applied bias are shown in Figs. 3(e) and 3(f), respectively. We observed a threshold behavior. When the applied bias is larger than the threshold voltage ($\sim$ 4 V), switchable hysteresis loop appears in PFM phase. In Fig. 3(e), coercive bias voltage of $\sim$ 4 V gives the threshold voltage to switch the charge polarization. A characteristic ``butterfly'' loop is also observed in PFM amplitude response. Bias voltage of the minima in PFM amplitude is the switch voltage in PFM phase. On the contrary, no hysteresis loops or butterfly loops are observed when the applied bias voltage is smaller than the threshold voltage. The existence of hysteresis loop together with threshold voltage was recognized as the solid evidence for ferroelectricity in traditional ferroelectric insulator, such as BaTiO$_3$\cite{BTO_AlO}, as well as in novel topological semimetal WTe$_2$\cite{WTe2}. Furthermore, to directly visualize the switchable polarization state, topograph and PFM images after poling are shown in Figs. 4(g)-(i). We applied negative (positive) voltage on the left (right) half of the scanning region. In Fig. 4(g), there is no contrast between left and right in the surface height. Both the PFM amplitude and phase images exhibit clear contrasts between left and right area due to the oppositely oriented remnant polarization after poling, which is also a typical feature of ferroelectricity. Therefore, our PFM measurements suggest TaNiTe$_5$ has a ferroelectric-like state on the surface and the polarization is switchable under the external bias.

At last, we would like to have some discussions about the interplay between the Dirac-like surface states and the novel surface reconstruction in TaNiTe$_5$. In our DFT calculations, we use the bulk crystal structure to calculate the surface state. In other words, no surface reconstruction is considered in our calculations. However, our calculations reproduce the observed surface states very well. The surface reconstruction seems to have very little effect on the surface band dispersions. This behavior can be understood to some extent. Taking surface band S$_1$ as an example, it exists in the bulk energy gap $\Delta_1$ (Fig. 2(i)). S$_1$ band above the Dirac point D$_1$ merges to bulk conduction band and S$_1$ band below the Dirac point merges to bulk valence band, which is in analogy to the surface band in topological insulator Bi$_2$Se$_3$\cite{BiSe}, implying that S$_1$ band has topological origin and fully determined by the bulk band. Since the surface reconstruction in TaNiTe$_5$ does not change the lattice periodicity, so no changes in the bulk bands occur, such as band folding, variation of bulk gap size or position, and so on. Therefore, the surface states are not altered by the surface reconstruction.

In summary, we have observed the coexistence of Dirac-like surface states and the switchable ferroelectric-like polarization on the (010) surface of the topological semimetal TaNiTe$_5$. The observed noncentrosymmetric surface reconstruction does not change the original surface lattice periodicity. Our STM can only determine the position of the surface Te atoms. High resolution TEM experiments revealing the position of other atoms below the surface Te atoms will help to understand the mechanism of the noncentrosymmetric reconstruction in the future. The observed surface states nicely match the DFT calculations, insensitive to the exotic surface reconstruction in this system. Our findings provide an interesting platform with ferroelectric-like charge distribution and robust Dirac-like surface states on the same surface.

This work was supported by the Ministry of Science and Technology of China (Grants No. 2016YFA0301003, 2016YFA0300501) and the National Natural Science Foundation of China (Grants No. U1632272, 11574201, 11521404, 11804194, U1632102, 11674224, 11974243, 11774223, U1732154, 11974061, 12004337) and the additional support from a Shanghai talent program. D. Q. acknowledges support from the Changjiang Scholars Program.


\end{document}